\documentclass[runningheads]{llncs}
\usepackage[T1]{fontenc}
\usepackage{graphicx}

\usepackage{url}            
\usepackage{booktabs}       
\usepackage{amsfonts}       
\usepackage{nicefrac}       
\usepackage{microtype}      
\usepackage{xcolor}         
\usepackage{multirow}
\usepackage{amsmath}
\usepackage{graphicx}
\usepackage{makecell}
\usepackage{verbatim}
\usepackage{caption}
\usepackage{subcaption}
\usepackage{booktabs}
\usepackage{latexsym}
\usepackage{comment}
\usepackage[misc]{ifsym}
\DeclareMathOperator*{\argmax}{arg\,max}
\usepackage{color}

\begin{document}
\title{Reformulating Sequential Recommendation: Learning Dynamic User Interest with Content-enriched Language Modeling}
\titlerunning{Language Modeling for Content-enriched Recommendation}
%
\author{Submitted for Blind Review}
\author{Junzhe Jiang \inst{1} \and
Shang Qu \inst{1} \and
Mingyue Cheng \inst{1} \Letter \and
Qi Liu \inst{1} \and
Zhiding Liu \inst{1} \and
Hao Zhang \inst{1} \and
Rujiao Zhang \inst{1} \and
Kai Zhang \inst{1} \and
Rui Li \inst{1} \and
Jiatong Li \inst{1} \and
Min Gao \inst{2}
}
\authorrunning{J. Jiang et al.}
%
\institute{State Key Laboratory of Cognitive Intelligence, University of Science and Technology of China, Hefei, China \and
The First Affiliated Hospital of University of Science and Technology of China, Hefei, China \\
\email{\{jzjiang,qushang\}@mail.ustc.edu.cn},
\email{\{mycheng,qiliuql\}@ustc.edu.cn}}
%
\maketitle              
\begin{abstract}
Recommender systems are indispensable in the realm of online applications, and sequential recommendation has enjoyed considerable prevalence due to its capacity to encapsulate the dynamic shifts in user interests. However, previous sequential modeling methods still have limitations in capturing contextual information. The primary reason is the lack of understanding of domain-specific knowledge and item-related textual content. Fortunately, the emergence of powerful language models has unlocked the potential to incorporate extensive world knowledge into recommendation algorithms, enabling them to go beyond simple item attributes and truly understand the world surrounding user preferences. To achieve this, we propose LANCER, which leverages the semantic understanding capabilities of pre-trained language models to generate personalized recommendations. Our approach bridges the gap between language models and recommender systems, resulting in more human-like recommendations. We demonstrate the effectiveness of our approach through a series of experiments conducted on multiple benchmark datasets, showing promising results and providing valuable insights into the influence of our model on sequential recommendation tasks. Furthermore, our experimental codes are publicly available at \url{https://github.com/Gnimixy/lancer}.

\end{abstract}
\section{Introduction}

Recommender systems are a crucial engine of various online applications, including e-commerce, advertising, and online videos. They play a pivotal role in discovering user interests and alleviating information overload, thus assisting users in their decision-making process. Over the past years, collaborative filtering (CF)~\cite{sarwar2001item,zhao2023cross,zhang2023towards} has become one of the most prevalent techniques due to its effectiveness and efficiency in large-scale recommendation scenarios.


By carefully revisiting recent progress in the sequential recommendation, we notice much of the great progress has been driven by achievements in network architecture or pre-training algorithms in natural language processing~\cite{de2021transformers4rec,zhang2022graph}. For one thing, a large body of advanced architecture models~\cite{kang2018self,liu2022one} has facilitated the building of high-capacity recommendation models. For another, some researchers attempt to leverage the powerful capacity of language modeling in semantic understanding to perform content-based sequential recommendations~\cite{wang2022transrec,cheng2022towards}. We also notice a few related advancements attempt to integrate pre-trained language models (PLMs) into recommender systems, leveraging their ability to capture semantic information and achieving promising results~\cite{liu2023pre,luo2023unlocking}. However, these works do not depart from the previous paradigm, merely adding a text encoder and continuing to use traditional sequential modeling methods.

With the continuous improvement of model capabilities, we ponder on whether sequential recommendation can be revolutionized to recommend items in a way that mimics human reasoning. Some recent works have attempted to use powerful PLMs as recommender systems by formulating recommendation as a language modeling task, with the item IDs~\cite{geng2022recommendation} or titles~\cite{zhang2021language} as inputs to the language model. Despite leveraging the inferential capabilities of language models for generating recommendations, these methods yielded unsatisfactory performance, falling behind the previous state-of-the-art results.

\begin{figure}[t]
  \centering
  \includegraphics[width=\linewidth]{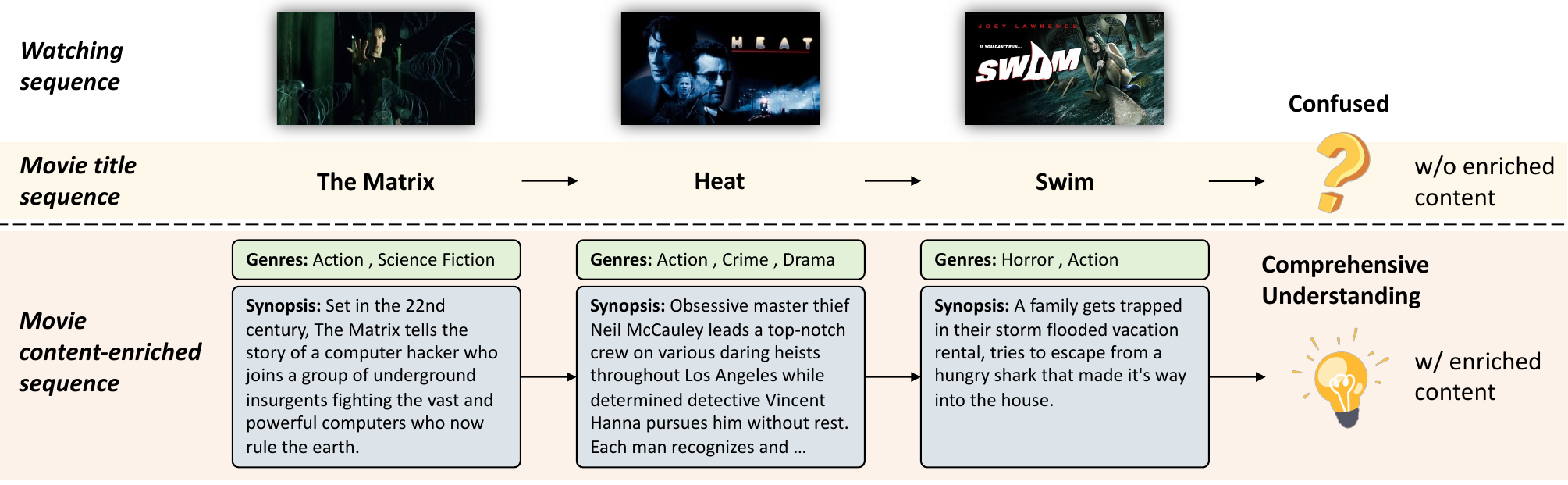}
  \caption{Comparison of sequence recommendation with and without content-enriched information. It can be seen that enriched content information provides more information about the content of the movies, which helps to reduce ambiguity and improve the accuracy of the recommendations.}
  \label{fig_paradigm_comparison}
\end{figure}

In light of the above, language models perform suboptimally to sequential recommendation systems due to their lack of domain-specific knowledge and mastery of item-related textual content. For example, in movie recommendation, \textit{The Matrix} could be understood by the language model as a mathematical matrix, rather than being associated with hacker-related information, which would significantly reduce the effectiveness of the recommendation. This limitation stems from the absence of side information in sequential recommendation, such as movie genres and synopses, which hinders a comprehensive understanding of user preferences underlying item choices (see Figure~\ref{fig_paradigm_comparison}).

To address these issues, we propose \textbf{LAN}guage modeling for \textbf{C}ontent-\textbf{E}nriched \textbf{R}ecommendation (LANCER) to employ PLMs as sequential recommenders by combining domain-specific knowledge and user behavior. Specifically, we introduce a \textbf{knowledge prompt} that facilitates the acquisition of domain-specific knowledge, and a \textbf{reasoning prompt} that integrates the acquired knowledge and item content information to generate personalized recommendations. To the best of our knowledge, we are the first to integrate enriched content into a language model-based sequential recommender system by transforming the sequential recommendation task into a text generation task.

The contributions of this paper are summarized as follows:

\begin{itemize}
    \item We propose LANCER, which incorporates domain knowledge and item content prompts into PLMs, leveraging their semantic understanding abilities to generate user recommendation results.
    \item Our designed model offers the advantage of integrating external knowledge to enhance the modeling of item content understanding. Our approach bridges the gap between language models and recommender systems by fusing abilities from both domains.
    \item We conduct extensive experiments on public datasets and demonstrate that our method achieves promising results, outperforming state-of-the-art methods in real-world recommendation scenarios. Furthermore, we provide valuable insights into the influence of our model on sequential recommendation tasks.
\end{itemize}

\section{Methodology}



\begin{figure}[t]
  \centering
  \includegraphics[width=1.0\linewidth]{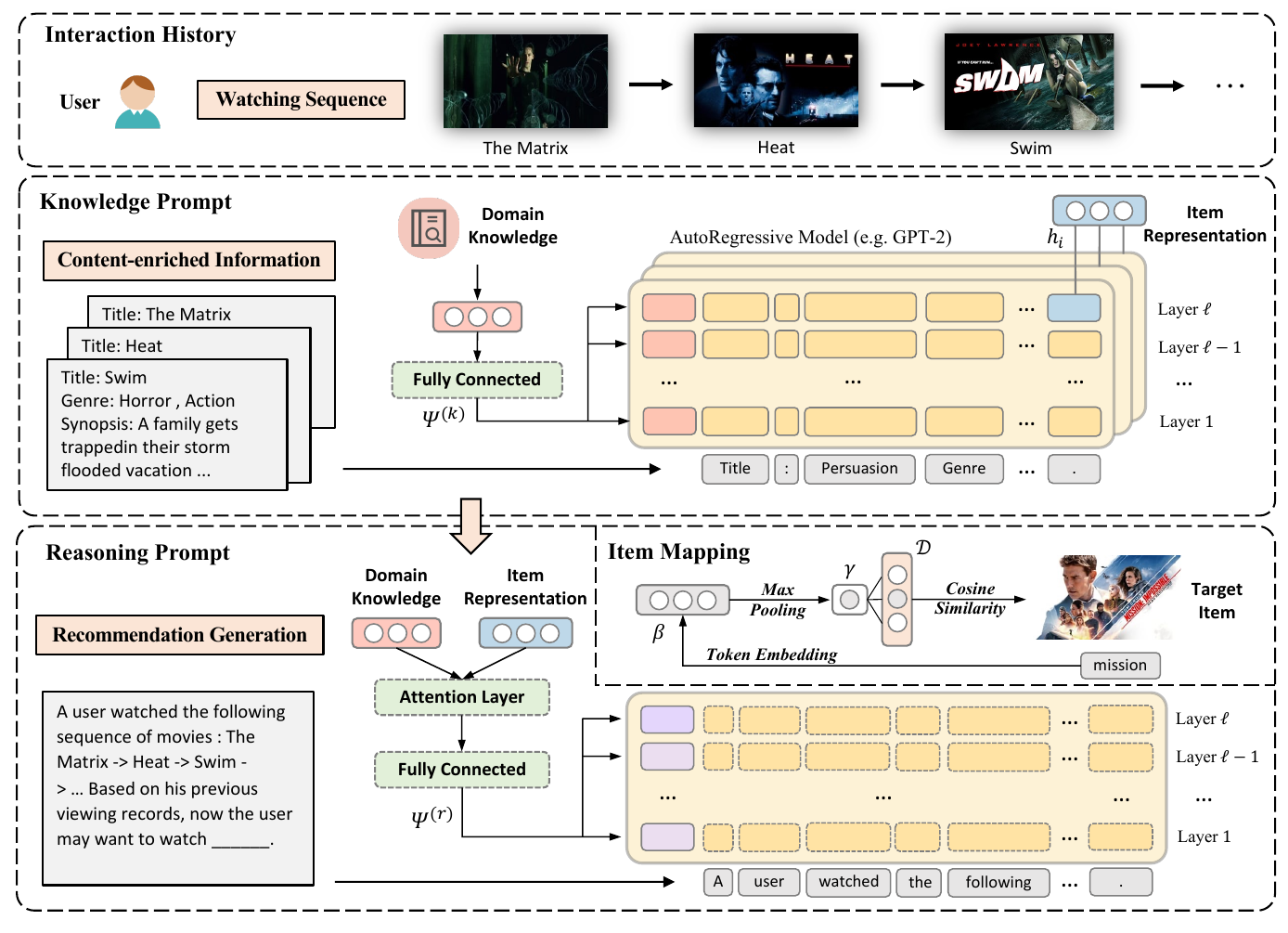}
  \caption{Illustration of the proposed LANCER, where the dashed line indicates the trainable parameters and the solid line indicates the frozen parameters.}
  \label{fig_architecture}
\end{figure}

\subsection{Model Architecture.} \label{ma}

The proposed LANCER architecture is shown in Figure \ref{fig_architecture}. For each item in the user interaction history, we extract all text content information and input it into a language model with fixed parameters to train the knowledge prompt and obtain a vector representation $h_i$ for each item. Afterward, we use an attention mechanism to combine domain knowledge with user interaction histories to form the reasoning prompt, allowing the model to generate the next item that the user may be interested in. Finally, the generated text is mapped to one of the candidate items to be recommended to the user.

\subsection{Knowledge Prompt.} \label{kp}

When directly applying a language model for item recommendation, a major challenge arises, as language models may not possess the knowledge of the relevant recommendation domains.

To address the above issues, we design continuous prompts to learn recommendation domain information while avoiding catastrophic forgetting that may occur with language model retraining. Inspired by Prefix-tuning~\cite{li2021prefix} and P-tuning~\cite{liu2022p}, we add continuous prefix prompts to guide the PLM without updating the original parameters. We define these as knowledge prompts.

We set a fixed number of knowledge tokens $\varepsilon^{(k)} = [e^{(k)}_1, ..., e^{(k)}_\theta]$, where $e^{(k)}_j$ represents the $j$-th knowledge token, and $\theta$ denotes the number of knowledge tokens. Using the following formula, we compute the knowledge prompt:

\begin{equation}
    \Psi^{(k)} = \text{MLP} \left( \text{Emb}(\varepsilon^{(k)}) \right),
\end{equation}

\noindent where $\text{MLP}$ denotes the multi-layer perceptron, $\text{Emb}$ denotes the embedding layer of the language model $\Theta_{f}$ with $L$ layers. During attention computation in the $\ell$-th layer $\ell \in \{1,2,...,L\}$, a distinct prefix of key-value pairs $Z_{\ell, K}$ and $Z_{\ell, V}$ is learnt. $ \Psi^{(k)}$ concatenates the parameters of the language model in each layer in the form of $\Psi^{(k)}_{\ell,K}$ and $\Psi^{(k)}_{\ell,V}$. The effect of the knowledge prompt propagates throughout the entire transformer activation layer and subsequently to the following tokens, enhancing them as follows:

\begin{equation}
    Z_{\ell, K}^{'} = [\Psi^{(k)}_{\ell,K} ; Z_{\ell, K}],\quad Z_{\ell, V}^{'} = [\Psi^{(k)}_{\ell,V} ; Z_{\ell, V}],
\end{equation}

\noindent where $Z_{\ell, K}^{'}, Z_{\ell, V}^{'} \in \mathbb R^{(\theta + M)\times d}$, $M$ denotes the input length of the language model, and $d$ denotes the dimension of hidden layers.
For each item $i$, we utilize the last hidden state $h_i$ obtained from the language model's last input token, considering all preceding tokens. Equation \ref{last_hidden_states} shows how the hidden layer is calculated:

\begin{equation}
    \label{last_hidden_states}
     h_i = \Theta_{f} (c_i \mid Z_{K}^{'}, Z_{V}^{'}),
\end{equation}

\noindent where $c_i$ denotes the detailed text information of item $i$. Therefore, a user's behavior sequence can be expressed as $H = [h_1,..., h_i,...,h_{n-1}]$. We will pad if the length of the sequence less than the fixed $N$.

At this stage, the training objective focuses on the detailed text information of items. The loss function for $\Theta_{f}$ is depicted in Equation~\ref{loss_function}, by minimizing the negative log-likelihood of predicting the next token, denoted as $e_j$. Training involves backpropagation to update the $\text{Emb}$ and $\text{MLP}$ corresponding to the knowledge prompts, while the parameters of $\Theta_{f}$ remain frozen.

\begin{equation}
    \label{loss_function}
    \mathcal{L} = \sum_j -\log p\left(e_n \mid e_{1},..., e_{j-1} ; \Theta_{LM}\right).
\end{equation}

\subsection{Reasoning Prompt.} \label{rp}

With the help of knowledge prompts, we can provide more accurate recommendations for a specific domain. However, this method lacks a modeling of user behavior and a deeper understanding of item information. Yet, user preferences often lie within a more personalized subset of domain information. 





To address this issue, we propose reasoning prompts, which employ an attention mechanism to combine domain knowledge and user behavior. In this process, the user's historical behavioral representation serves as the query, while the domain-related information acts as the key and value. 
Then we extract the unique reasoning prompt $\Psi^{(r)}$ for each user to help the subsequent model to better provide recommendations, where $A$ is the result calculated by the attention mechanism:

\begin{gather}
    \label{reasoning_prompt}
    \Psi^{(r)} = \sigma \left( W_r \times A + b_r \right).
\end{gather}

In the generation phase, we use beam search decoding to generate output sequences with the $w$ highest scores, then truncate each output to include only the first new item. Beam search decoding at step $t$ can be expressed as:

\begin{gather}
    \pi_t = \{[x_{t-1}, x'_t] \ | \ x'_t \in \Omega, x_{t-1} \in \pi_{t-1} \}, \\
    \alpha_t = \argmax_{\lambda \in \pi_t, |\lambda|=w} \log P(\lambda \mid \Theta(T, \Psi^{(r)})),
\end{gather}

\noindent where $\Omega$ is the vocabulary set of trainable language model $\Theta$, $w$ is the number of beams, $e'_t$ denotes a particular text generated at the $t$-th step during beam search decoding, and $\lambda$ is the current set of top-$w$ outputs. 

During this phase, the training objective is to enable the language model $\Theta$ to predict text that corresponds to the textual information of the item the user will choose next.

\subsection{Item Mapping.} \label{im}

At this point, we have obtained the results of the model prediction, but we still need to map the predicted result to one of all items in the database $\mathcal{D}$. This matching process is conducted as follows: initially, we derive the vector representation sequence of the predicted item $n$, denoted as $\beta = \text{Emb}(\alpha^{(n)})$, where $\beta \in \mathbb R^{|\beta| \times d}$ and $|\beta|$ represents the sequence length of $\beta$. Subsequently, we calculate the maximum value for each dimension of the $\beta$ sequence to represent the predicted item as a fixed-length vector, where the maximum value $\gamma_j$ for the $j$-th dimension is:

\begin{equation}
    \label{max_pooling}
    \gamma_j = \max (\beta_j), \quad \text{for}\ j=1,2,...,|\beta|.
\end{equation}

Finally, we compute the cosine similarity ($\textit{CosSim}$) in equation \ref{cosine_similarity} between the vector representations of the output $\gamma$ and the titles of each item in the database $\mathcal{D}$, then identify the item with the highest similarity. The representations of all items can be computed only once and used throughout the generation process.

\begin{equation}
    \label{cosine_similarity}
    s_n = \mathop{\arg\max}_{\delta \in \mathcal{D}} \textit{CosSim}(\gamma, \delta).
\end{equation}

\section{Experiments}


\subsection{Experimental Setup}

\paragraph{Datasets.}

We experimented with three datasets: MovieLens\footnote{\url{https://grouplens.org/datasets/movielens/}}, MIND\footnote{\url{https://msnews.github.io/}}~\cite{wu2020mind}, and Goodreads\footnote{\url{https://sites.google.com/eng.ucsd.edu/ucsdbookgraph/home}}~\cite{wan2018item}. The detailed statistics of the datasets are presented in Table \ref{statistics}. MovieLens is for movie recommendations, MIND for English news, and Goodreads for books reviews. We organized user data chronologically and limited sequence lengths to 5-10 items, excluding shorter sequences and truncating longer ones.

\paragraph{Evaluation Metrics.}
To evaluate our model, we compute Recall and NDCG for the top 5 and 10 items respectively. The former is an evaluation of un-ranked retrieval sets while the latter reflects the order of ranked lists. We follow the leave one-out strategy described in~\cite{kang2018self} to split the interactions into training, validation, and testing sets. For the testing stage, we adopt the all-ranking protocol to evaluate recommendation performance.

\begin{table}[t] 
    \caption{Statistics of the datasets after processing.}
    \label{statistics}
    \centering
    \renewcommand\arraystretch{1.1}
    \resizebox{\linewidth}{!}{
    \setlength{\tabcolsep}{1mm}{
    \begin{tabular}{ccccccc}
    \toprule
        Dataset & Num. User & Num. Item & Num. Inteaction & Sparsity & Avg. User & Avg. Item \\
    \midrule
        MovieLens & 6,040 & 3,231 & 72,480 & 99.63\% & 12.00  & 22.43 \\
        MIND & 91,935 & 44,908 & 2,248,027 & 99.94\% & 24.45 & 50.06 \\ 
        Goodreads & 120,968 & 28,480 & 1,095,758 & 99.97\% & 9.06 & 38.47 \\
    \bottomrule
    \end{tabular}
    }}
\end{table}

\paragraph{Implementation Details.}

We use the GPT-2~\cite{radford2019language} pre-trained model from Hugging Face \footnote{\url{https://huggingface.co/gpt2}} as our backbone model for experiments. During training, we employ the Adam optimizer with a linear schedule learning rate decay. For different datasets, we keep the following hyperparameters consistent: batch size of 16, maximum token length of 512 for the language model input, maximum input length of 32 tokens for each item (truncated for items exceeding the limit). We employ the Sequential Tuning approach from~\cite{li2023personalized}.
Our model is trained on NVIDIA Tesla A100 GPUs with 40GB of memory.

\paragraph{Compared Methods.}

We compare the performance of our model to various baselines, including ID-based and content-based methods. In addition, we evaluate the performance of recent methods which also use PLMs as the main model architecture. Specifically, we compare our model to the following baselines, where Caser~\cite{tang2018personalized}, GRU4Rec~\cite{hidasi2015session}, SASRec~\cite{kang2018self}, BERT4Rec~\cite{sun2019bert4rec}, S\textsuperscript{3}Rec~\cite{zhou2020s3}, CL4SRec~\cite{xie2022contrastive} are ID-based,  and , FDSA~\cite{zhang2019feature}, MV-RNN~\cite{cui2018mv}, SASRec+~\cite{kang2018self} are content-based.

\begin{table}[t]
    \centering
    \caption{Recommendation performance comparison of different recommendation models over three datasets, in which the best results are in bold and the second best results are underlined. R and N denote Recall and NDCG respectively.}
    \renewcommand\arraystretch{1.2}
    \resizebox{\linewidth}{!}{
    \setlength{\tabcolsep}{1mm}{
    \begin{tabular}{ccccccccccccc}
    \toprule
        \multirow{2}{*}{Model} & \multicolumn{4}{c}{MovieLens} & \multicolumn{4}{c}{MIND} & \multicolumn{4}{c}{Goodreads} \\ \cmidrule(r){2-13}
        ~ & R@5 & R@10 & N@5 & N@10 & R@5 & R@10 & N@5 & N@10 & R@5 & R@10 & N@5 & N@10 \\
    \midrule
        Caser & 0.0225 & 0.0330 & 0.0151 & 0.0185 & 0.0312 & 0.0575 & 0.0181 & 0.0265 & 0.1058 & 0.1452 & 0.0770 & 0.0898 \\
        GRU4Rec & 0.0737 & 0.1316 & 0.0450 & 0.0635 & 0.0981 & 0.1589 & 0.0628 & 0.0816 & 0.1871 & 0.2472 & 0.1394 & 0.1587 \\
        SASRec & 0.0967 & \underline{0.1589} & 0.0624 & 0.0823 & 0.1087 & 0.1707 & 0.0714 & 0.0913 & 0.2043 & 0.2641 & \underline{0.1536} & 0.1729 \\
        BERT4Rec & 0.0776 & 0.1391 & 0.0466 & 0.0664 & 0.0853 & 0.1359 & 0.0544 & 0.0706 & 0.1892 & 0.2487 & 0.1391 & 0.1583 \\
        S\textsuperscript{3}Rec & 0.0821 & 0.1469 & 0.0509 & 0.0716 & 0.0990 & 0.1586 & 0.0639 & 0.0830 & 0.1753 & 0.2478 & 0.1207 & 0.1442 \\
        CL4SRec & \underline{0.0996} & 0.1527 & \underline{0.0683} & \underline{0.0853} & 0.0964 & 0.1543 & 0.0625 & 0.0811 & 0.1986 & 0.2599 & 0.1477 & 0.1675  \\
    \midrule
        FDSA & 0.0901 & 0.1407 & 0.0595 & 0.0757 & 0.1063 & \underline{0.1728} & 0.0682 & 0.0896 & 0.1713 & 0.2351 & 0.1195 & 0.1400 \\
        MV-RNN & 0.0853 & 0.1399 & 0.0527 & 0.0703 & 0.1024 & 0.1619 & 0.0656 & 0.0847 & 0.1872 & 0.2486 & 0.1385 & 0.1583 \\
        SASRec+ & 0.0944 & 0.1454 & 0.0618 & 0.0781 & \underline{0.1107} & \textbf{0.1753} & \underline{0.0723} & \textbf{0.0930} & \underline{0.2045} & \underline{0.2663} & 0.1533 & \underline{0.1733} \\
    \midrule
        LANCER & \textbf{0.1086} & \textbf{0.1616} & \textbf{0.0712} & \textbf{0.0884} & \textbf{0.1168} & 0.1675 & \textbf{0.0766} & \textbf{0.0930} & \textbf{0.2121} & \textbf{0.2673} & \textbf{0.1600} & \textbf{0.1779} \\
    \bottomrule
    \end{tabular}}}
    \label{main}
\end{table}

\subsection{Results and Analyses}
\paragraph{Recommendation Performance Evaluation.}

In terms of sequential recommendation, our experimental results in Table \ref{main} demonstrate that LANCER outperforms previous models across movie, news, and book recommendations. Therefore, it can be seen that LANCER can integrate various textual features of items well, utilizing the powerful text understanding and generation capabilities of PLMs. Additionally, LANCER incorporates knowledge and item features in the recommendation domain, delivering more personalized and precise recommendations to users.

Additionally, we notice observe several notable points in our results. First, LANCER performs more prominently on MovieLens, which we attribute to two factors: 1) the relatively small size of the dataset allows our model to leverage the extensive knowledge contained in the pre-trained model, unleashing its full potential. 2) Movie titles offer limited information, and further integration of details like genre, synopsis, director, etc., along with user movie-watching behavior, could enhance reasoning and recommendations.
Second, we notice that the improvement in the @5 metrics is greater than that in the @10 metrics, which we attribute to the generation capabilities of the language model. Generative LMs often perform better in the first few generations, but the quality of generations tends to decline as the number of generations increases. The reason we choose GPT-2 as our backbone is that generative autoregressive models better fit the characteristics of sequential data.

\paragraph{Case Study.}

\begin{figure}[t]
  \centering
  \includegraphics[width=\linewidth]{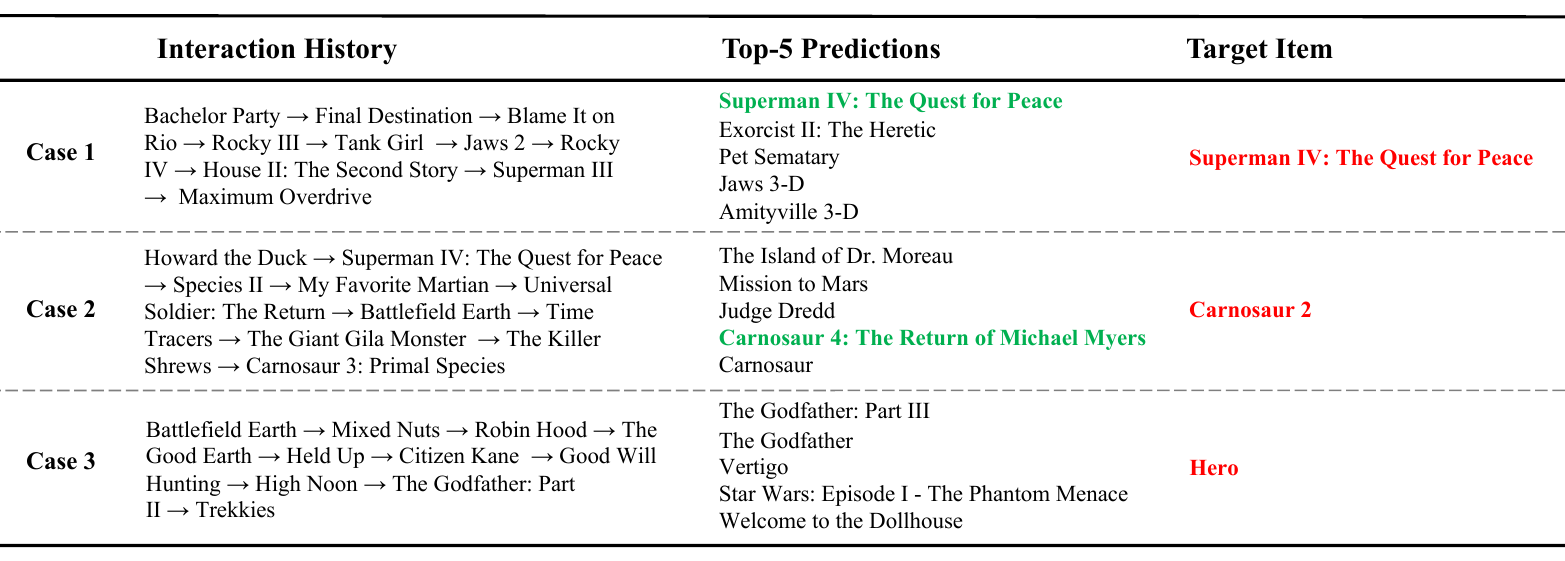}
  \caption{Case study of three users from the MovieLens dataset in terms of top-5 ranked predictions. The texts in green indicate predictions that map to the target.}
  \label{fig5}
\end{figure}

In Figure~\ref{fig5}, we select three user examples from the MovieLens-1M dataset and present the top 5 output texts directly generated by the proposed LANCER. It is evident that our model is capable of generating recommendations that align reasonably well with user preferences. Furthermore, we observe that model outputs are realistic movie titles in almost all scenarios, even when the model fails to accurately recommend the next item. Case 2 is among the few examples where the model output do not exactly match a movie title in the database. In this case, the original model output is mapped to \textit{Carnosaur 2}, which is the correct next item. In Case 3, although the target item does not appear in the top 5 predictions, several predicted movies share overlapping genres with the target. Both \textit{Welcome to the Dollhouse} and \textit{Hero} are \textit{Comedy|Drama} movies, and \textit{The Godfather} also lies in the \textit{Drama} category.

\section{Conclusion}

In conclusion, we propose LANCER, which utilizes language models as sequential recommender systems, effectively integrating domain-specific knowledge and general knowledge of the PLM to learn dynamic user interests. Our approach employs a knowledge prompt to facilitate the acquisition of item content and a reasoning prompt to enable the integration of both domain-specific knowledge and user behavior for personalized recommendation. We highlight the potential of our approach in bridging the gap between language models and recommender systems and provide experimental results to support its effectiveness in incorporating item content understanding.


\subsubsection{Acknowledgements.} This research was supported by grants from the National Natural Science Foundation of China (Grants No. 62337001, 623B1020) and the Fundamental Research Funds for the Central Universities, the Fundamental Research Funds for the Central Universities and the Joint Fund for Medical Artificial Intelligence (Grant No. MAI2022C004).
%
%
%
%





\bibliographystyle{splncs04}
\bibliography{references}

\end{document}